\documentclass[aps,prb,twocolumn,superscriptaddress,nofootinbib]{revtex4-2}

\usepackage{graphicx,url,hyperref,dcolumn,bm,multirow,amsmath,amssymb,amsfonts,subfigure,soul,wrapfig}
\usepackage[dvipsnames,svgnames,x11names]{xcolor}
\usepackage[symbol]{footmisc}
\usepackage[normalem]{ulem}
\usepackage[version=4]{mhchem}
\usepackage{placeins}

\hypersetup{colorlinks,linkcolor=Red,urlcolor=Emerald,citecolor=Blue}
\usepackage{float}
\usepackage[scaled]{helvet}
\usepackage[T1]{fontenc}

\begin{document}
	
	\title{Thermodynamic signatures of a field-induced ordered intermediate phase in \ce{Na2Co2TeO6}}
	
\author{Prashanta K. Mukharjee$^{*\,\dagger}$}
\affiliation{Experimental Physics VI, Center for Electronic Correlations and Magnetism, Institute of Physics, University of Augsburg, 86159 Augsburg, Germany}

\author{Sebastian Erdmann$^{*}$}
\affiliation{Experimental Physics VI, Center for Electronic Correlations and Magnetism, Institute of Physics, University of Augsburg, 86159 Augsburg, Germany}

\author{R. Kalaivanan}
\affiliation{Institute of Physics, Academia Sinica, Taipei 11529, Taiwan}

\author{R. Sankar}
\affiliation{Institute of Physics, Academia Sinica, Taipei 11529, Taiwan}

\author{Kwang-Yong Choi}
\affiliation{Department of Physics, Sungkyunkwan University, Suwon 16419, Republic of Korea}

\author{Alexander A. Tsirlin}
\affiliation{Felix Bloch Institute for Solid State Physics, University of Leipzig, 04103 Leipzig, Germany}

\author{Philipp Gegenwart$^{\ddagger}$}
\affiliation{Experimental Physics VI, Center for Electronic Correlations and Magnetism, Institute of Physics, University of Augsburg, 86159 Augsburg, Germany}
	
	\date{\today}
	
	\maketitle
	
\begingroup
\renewcommand{\thefootnote}{\fnsymbol{footnote}}
\footnotetext[1]{These authors contributed equally to this work.}
\footnotetext[2]{Contact author: pkmukharjee92@gmail.com}
\footnotetext[3]{Contact author: philipp.gegenwart@physik.uni-augsburg.de}
\endgroup
	\newpage
	
	\begin{center}
		\textbf{ABSTRACT}
	\end{center}
	
	\textbf{The honeycomb cobaltate \ce{Na2Co2TeO6} has recently been proposed as a candidate material for hosting field-induced quantum spin liquid (QSL) behavior. Here, we present a comprehensive thermodynamic study of its low-temperature, high-field phase diagram using magnetization, specific heat, and magnetocaloric-effect measurements down to 1 K.  In zero field, we observe a weak residual moment that provides further insight into the nature of the magnetic ground state. For in-plane magnetic fields ($B \parallel a^*$), we identify three field-induced transitions at $B_{c1} \simeq 6$ T, $B_{c2} \simeq 7.8$ T, and $B_{c3} \simeq 10.4$ T. The magnetic Grüneisen parameter and specific heat reveal clear thermodynamic signatures of these successive phase transitions enclosing two intermediate phases. Contrary to expectations for a field-induced QSL, the phase between $B_{c2}$ and $B_{c3}$ lacks enhanced magnetic entropy but instead shows behavior consistent with a distinct ordered state. Above $B_{c3}$, the absence of additional anomalies indicates a crossover to a conventional spin-polarized regime. Our results place stringent thermodynamic constraints on the proposed QSL scenario in \ce{Na2Co2TeO6}, calling for further microscopic investigations to establish the precise nature of the field-induced phases.}
	
	\begin{center}
	
	\end{center}
		Quantum spin liquids (QSLs) are unconventional magnetic states characterized by long-range entanglement and fractionalized excitations~\cite{Broholm2020}. The Kitaev model with bond-dependent Ising interactions between spin-1/2 moments on a honeycomb lattice provides an exact theoretical realization of a unique QSL state~\cite{Kitaev20062}. Experimental realization of the Kitaev QSL remains challenging, as candidate honeycomb materials such as $\alpha$-RuCl$_3$ and (Li, Na)Ir$_2$O$_3$ typically develop magnetic long-range order (LRO) due to non-Kitaev interaction terms~\cite{Winter_2017}. Suppressing this order via magnetic field or pressure to reach a proximate Kitaev regime has become a common experimental strategy~\cite{Matsuda045003}. In this context, $3d^7$ cobalt-based honeycomb materials (Co$^{2+}$, \mbox{$S_{\mathrm{eff}}=\tfrac{1}{2}$}) have been proposed as promising alternatives to first-generation $4d/5d$ systems, as Kitaev interactions may become dominant in the cobaltates~\cite{Liu047201,Liu014407}. This has renewed interest in cobalt-based honeycomb compounds, among which \ce{Na2Co2TeO6} (NCTO) is particularly interesting due to its strong magnetic frustration, experimentally accessible field-tunable behaviour, and magnetic excitations that closely link to theoretical predictions of extended Kitaev–Heisenberg models~\cite{Lefrancois214416,Lin2021,Songvilay224429,Yao147202,Kim_2022}.
		
		In the archetypal Kitaev material $\alpha$-RuCl$_3$, an in-plane magnetic field suppresses antiferromagnetic LRO and drives the system into a high-field disordered regime whose microscopic nature as a possible QSL state remains vividly debated~\cite{Yokoi568,Kasahara2018,Bachus097203,Lefrancois2022,Ponomaryov037202,Banerjee2018}, with further studies still ongoing ~\cite{Sarkis2026}. Within this broader context, NCTO offers a complementary platform to test field-induced phenomena in honeycomb magnets. NCTO crystallizes in a layered hexagonal structure (space group $P6_3 22$), composed of honeycomb layers of edge-sharing CoO$_6$ octahedra separated by disordered Na layers along the $c$ axis~\cite{Bera094424,Xiao2019,Lee134411}. In zero magnetic field, NCTO exhibits antiferromagnetic LRO ($T_{N1} \simeq 28$~K), alongside two weaker anomalies at ~17 K and ~5 K that are suppressed by an in-plane magnetic field of approximately 8~T~\cite{Lin2021}. 
		 	\begin{figure*}[!ht]
			\centering
			\includegraphics[width=0.75\textwidth]{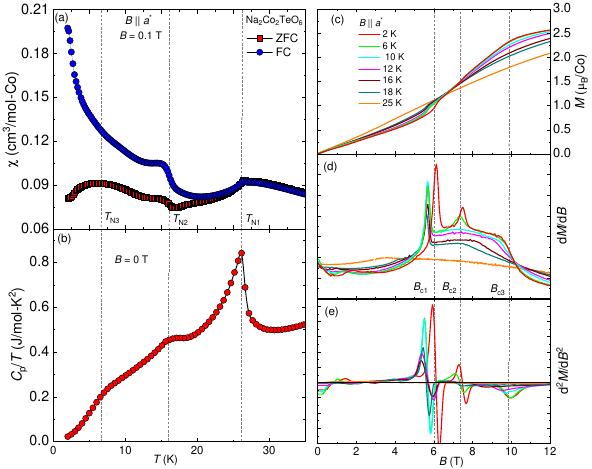}
			\caption{\sffamily\textbf{Magnetization and specific heat}: {\bf (a)} Zero-field cooled (ZFC) and field-cooled (FC) DC magnetic susceptibility versus $T$ measured at $B =0.1$~T. {\bf (b)} Specific heat of NCTO in $B$ = 0 T. The grey dotted lines indicate phase transitions. {\bf (c)} Field-dependent magnetization, $M(B)$, for NCTO at different temperatures down to 2\,K. {\bf (d,e)} The first and second derivative of $M(B)$. The grey dotted lines indicate the positions of the critical fields.}
			\label{Fig1}
		\end{figure*}
		Magnetic phase diagrams have been established for in-plane ($B \parallel a$ and $B \parallel a^*$)~\cite{ArnethL140402,Lin2021,Kikuchi224416,Zhang064421,Fang106701,TakedaL042035,Hong144426} and out-of-plane ($B \parallel c$)~\cite{Zhang144431,Arneth094001,Zhou241108} field orientations. 
		
		For $B \parallel a^*$, three field-induced anomalies, $B_{c1}$, $B_{c2}$, and $B_{c3}$, emerge within the magnetically ordered state, dividing the $B-T$ phase diagram into four distinct regions (phases I-IV)~\cite{Zhang064421,Fang106701,Hong144426}. In particular, the phase between $B_{c2}$ and $B_{c3}$ (phase III), as well as the high-field regime above $B_{c3}$ (phase IV), have attracted considerable attention, and the nature of these states remains under active debate. In torque magnetometry at 2~K, the angular pattern evolves from twofold ($C_2$) in phase I to nearly sixfold in phase II and becomes essentially sixfold ($C_6$) in phase III. Together with the broad excitation continuum observed in neutron scattering, this was interpreted as evidence for a field-induced QSL in phase III~\cite{Lin2024}. By contrast, angle-resolved specific-heat measurements rule out a Kitaev QSL in this field range, because the excitations remain gapped and the angular anisotropy of the gap is not consistent with the angle dependence expected for a Majorana gap in the Kitaev model~\cite{Fang106701}. Further, thermal transport studies in NCTO reveal a very small residual longitudinal thermal conductivity in zero magnetic field~\cite{Hong2024,Guang184423}. Under magnetic fields, a finite thermal Hall response ($\kappa_{xy}$) has been observed in both standard ($B \parallel c$)~\cite{Yang081116,Li140402,Gillig043110} and planar ($B \parallel a$ and $B \parallel a^*$)~\cite{TakedaL042035,Guang024419,Chen2024} configurations. On the one hand, such a finite $\kappa_{xy}$ may be expected in a field-induced QSL phase due to the emergence of topological magnetic excitation~\cite{Katsura066403}. On the other hand, $\kappa_{xy}$ was shown to persist far above $T_{N1}$~\cite{Yang081116}. It also reveals strong similarities to $\kappa_{xx}$, which is suggestive of the predominantly phononic origin~\cite{Chen2024,Hong094403,Yan2025}. Taken together, these experimental reports show that the nature of both phases III and IV is still unresolved, calling for a comprehensive thermodynamic investigation of NCTO at low temperatures and high magnetic fields.\\

	Here, we present a comprehensive thermodynamic study of NCTO single crystals using magnetization, specific heat, and magnetocaloric-effect measurements. In zero field, the system exhibits a series of transition anomalies, while the application of an in-plane magnetic field (for $B \parallel a^*$) gives rise to a rich field–temperature phase diagram with multiple field-induced transitions, consistent with the earlier reports~\cite{Zhang064421,Lin2021}. Focusing on the low-temperature, high-field regime, our multiple thermodynamic probes reveal clear signatures of successive phase transitions above $B_{c2}$, including distinct anomalies in the magnetic Gr\"uneisen parameter and specific heat. Notably, the thermodynamic response of phase III is inconsistent with expectations for a field-induced QSL.
Furthermore, the absence of additional thermodynamic anomalies at higher fields suggests a crossover into a conventional high-field regime. Our results provide a thermodynamic framework for understanding the field-induced phases in NCTO and offer important insights into the ongoing debate on their microscopic nature.\\
		
	\noindent \textbf{Magnetization and zero-field specific heat}\\
	Upon cooling, NCTO exhibits three magnetic transitions in zero field at $T_{N1}\simeq 28$~K, $T_{N2}\simeq 17$~K, and $T_{N3}\simeq 5$~K. As shown in Fig.~\ref{Fig1}(b), these appear in the zero-field specific heat as a pronounced sharp peak at $T_{N1}$, a weaker anomaly at $T_{N2}$, and a broad feature around $T_{N3}$. Consistently, the low-field temperature-dependent DC susceptibility [Fig.~\ref{Fig1}(a)] shows a clear anomaly at $T_{N1}$, and broad features around $T_{N2}$ and $T_{N3}$. The zero-field transition at $T_{N1}$ has been documented using various experimental probes and corresponds to the onset of antiferromagnetic long-range order~\cite{ArnethL140402,TakedaL042035,Lin2021,YaoL022045,Jiao159,Bera094424,Zhang064421}. On the other hand, the comparatively weaker anomalies near 17~K and 5~K have been attributed to effects associated with interlayer Na-ion disorder, domain reorientation and spin canting~\cite{Lin2021,Bera094424,Jiao159}. 
	
	The transition temperatures $T_N$ are determined from the Fisher's heat capacity [$\frac{d(\chi T)}{dT}$], as shown in Fig.~S1. With increasing applied magnetic field, the sharp peak associated with $T_{N1} \simeq 28$~K shifts progressively to lower temperatures. Above 4~T, this peak begins to broaden noticeably, and by 8~T, only a subtle change in slope remains visible in $\frac{d(\chi T)}{dT}$, indicating a significant suppression of the transition. On the other hand, the anomalies associated with $T_{N2}$ and $T_{N3}$ weaken progressively with increasing field. Above 2~T, any features corresponding to these two transitions become difficult to resolve, though a broad minimum remains discernible in $\chi(T)$ for fields in the range 3--5~T. A similar field-dependent behavior in magnetic susceptibility has been reported in previous studies~\cite{Zhang064421,Yao085120}. Given the weak nature of the $T_{N2}$ and $T_{N3}$ anomalies under an applied field, these transitions are not included in the $B-T$ phase diagram presented in Fig.~\ref{Fig5}.
			 
	\begin{figure}[!ht]
		\centering
		\includegraphics[width=1\linewidth]{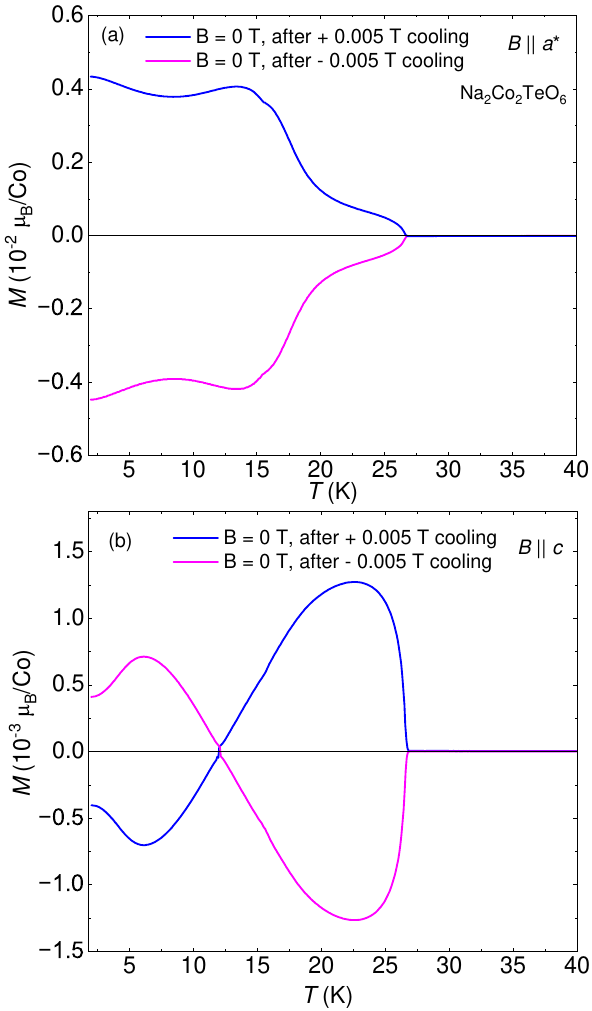}
		\caption{\sffamily\textbf{Residual moment}: Temperature dependent magnetization measured in zero field after cooling the sample in weak training fields for (a) in-plane ($B \parallel a^*$) and (b) out-of-plane ($B \parallel c$) directions.}
		\label{Fig2}
	\end{figure}
	
	To characterize the zero-field ground state, we also performed
	zero-field $M(T)$ measurements after cooling the sample in weak positive and negative
	training fields (Fig.~\ref{Fig2}), together with low-field $M(B)$ loops at 2~K (Fig.~S2). The zero-field temperature sweeps reveal a weak residual
	moment in the ordered state for both in-plane ($B \parallel a^*$) and out-of-plane ($B \parallel c$) training-field directions. For $B \parallel c$, the trained moment changes sign near 13~K and vanishes on approaching $T_N$, closely resembling the ferrimagnetic behavior reported previously~\cite{Yao085120}, although the absolute value of the uncompensated moment along the $c$ axis is about 5 times smaller compared to the earlier publication.
	Interestingly, we also observe a small ferromagnetic moment along $a^*$,
	which was not reported previously. Consistently, the low-field $M(B)$
	curves at 2~K show small but finite remanent moments and confirm weak
	ferromagnetism. Whereas weak ferromagnetism along $c$ is a natural
	consequence of the triple-$q$ state reported for NCTO~\cite{Francini235118,Francini075104,Krugger146702}, the presence of the in-plane ferromagnetic moment may indicate a deformation of this state and requires further dedicated investigation.
	
	\begin{figure*}[t]
		\centering
		\includegraphics[width=0.75\textwidth]{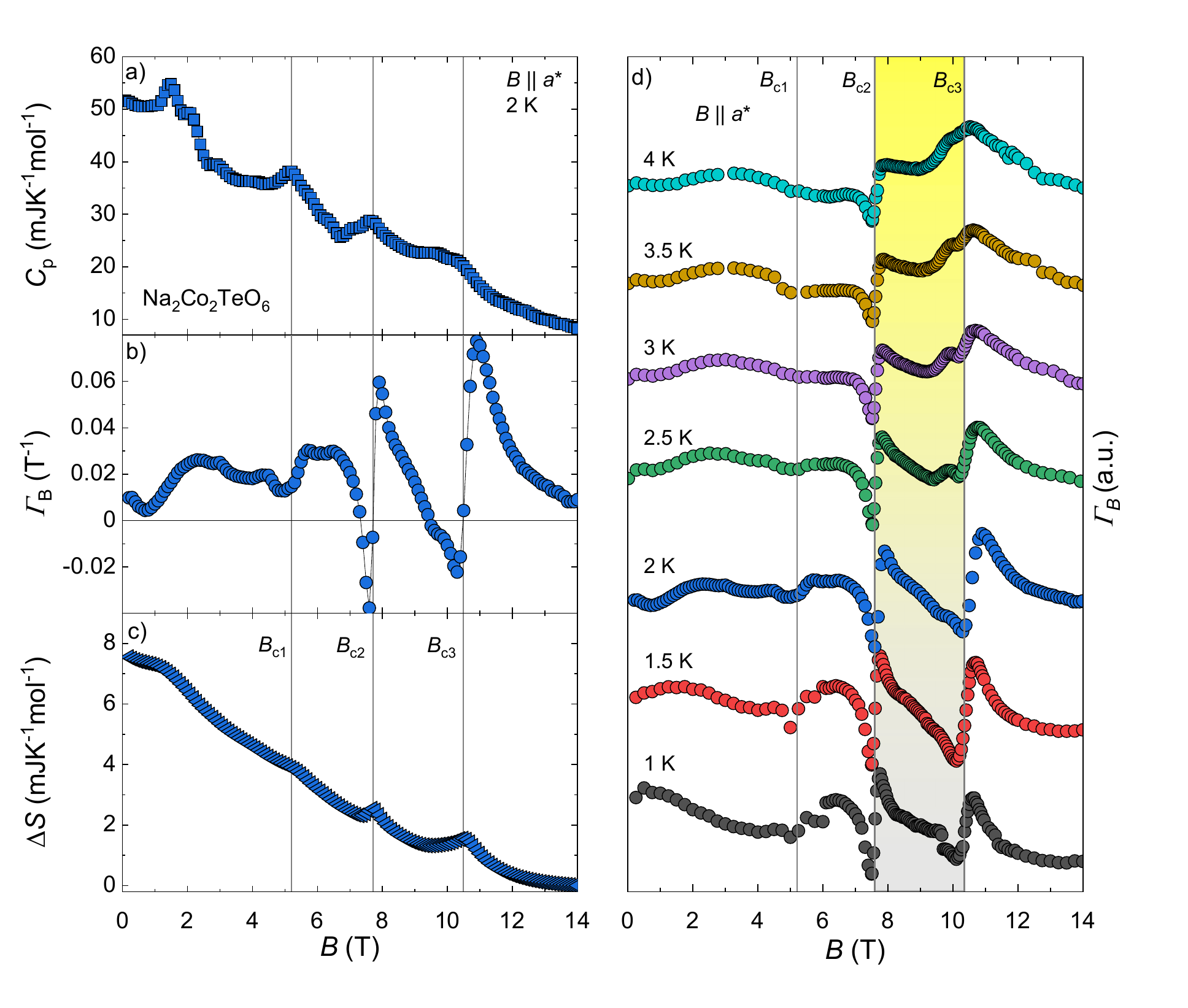}
		\caption{\sffamily\textbf{Field dependence of $\Gamma_B$ and specific heat}:
			{\bf (a)} Specific heat as a function of magnetic field at $T = 2$~K.
			{\bf (b)} Magnetic Gr\"uneisen parameter $\Gamma_B$ as a function of magnetic field at $T = 2$~K.
			{\bf (c)} Field evolution of magnetic entropy at $T = 2$~K.
			{\bf (d)} Field evolution of $\Gamma_B$ at different temperatures down to 1~K. The data are vertically offset for clarity. 
			The gray lines indicate the position of the critical fields.
			The region between $B_{c2}$ and $B_{c3}$ is presented in yellowish-gray color. The data in this figure are shown for the increasing field-sweep cycle.}
		\label{Fig3}
	\end{figure*}
	
   	In the following, we focus on the field-induced transitions in NCTO. Figure~\ref{Fig1}(c) shows the field-dependent magnetization $M(B)$ of NCTO measured over a broad temperature range below 28~K. Deep within the ordered state ($T=2$~K), the magnetization increases with field and reaches about $2.5~\mu_{B}$/Co at 12~T. The $M(B)$ curves display three field-induced anomalies at $B_{c1}$, $B_{c2}$, and $B_{c3}$, which are more clearly resolved in the first and second derivatives. In the first derivative $dM/dB$, $B_{c1}$ and $B_{c2}$ appear as sharp peaks, while $B_{c3}$ is best identified as a broad minimum in the second derivative $d^{2}M/dB^{2}$ [Fig.~\ref{Fig1}(d,e)]. Similar signatures have been observed in the field-dependent magnetization data in previous reports~\cite{Zhang064421,Lin2021}. The increasing $\leftrightarrow$ decreasing field-sweep curves exhibit a clear hysteresis at $B_{c1}$ up to temperatures of approximately 18~K, indicating the first-order nature of this transition~(see Fig.~S3). In contrast, no such hysteresis is observed at $B_{c2}$ and $B_{c3}$, suggesting that these transitions are of second-order character. 
		
	The evolution of these critical fields with temperature reveals distinct behaviors for each transition. $B_{c1}$, identified as a peak in $dM/dB$, shifts to lower fields with increasing temperature up to $\sim 12$~K, beyond which it remains nearly unchanged up to 20~K. The sharp peak at $B_{c2}$ in $dM/dB$ broadens into a wide maximum above 6~K, while its position shifts only weakly with temperature. The anomaly at $B_{c3}$, visible as a minimum in $d^{2}M/dB^{2}$, becomes progressively weaker with increasing temperature, and its center shifts smoothly to lower fields.

	The positions of all three critical fields at their corresponding temperatures are plotted in the $B-T$ phase diagram (see Fig.~\ref{Fig5}). In addition to the three distinct field-induced anomalies, the $dM/dB$ curves exhibits a broad low-field minimum around $1$--$4$~T, which is most clearly resolved at low temperatures. Corroborating signatures of this feature are also present in the specific heat and the magnetic Gr\"{u}neisen parameter $\Gamma_B$ (discussed in the next section).\\

	\noindent \textbf{Field-dependent thermodynamics} \\
		The field dependence of the specific heat and magnetic Gr\"uneisen parameter provides direct thermodynamic access to the phase evolution of NCTO. As shown in Fig.~\ref{Fig3}(a), the specific heat $C_p(B)$ at $ T =2$~K exhibits several anomalies upon increasing magnetic field. In the low-field region, $C_p(B)$ first shows a broad maximum accompanied by additional step-like features. More pronounced anomalies appear as clear maxima at $B_{c1} \simeq 5.8$~T and $B_{c2} \simeq 7.8$~T in the form of clear maxima, while a very broad maximum is observed at $B_{c3} \simeq 10.4$~T. Further insight is provided by the magnetic Gr\"uneisen parameter $\Gamma_B$. For brevity, we focus on the $\Gamma_B(B)$ curve at 2~K, as shown in Fig.~\ref{Fig3}(b). In the low-field regime below about 7~T, $\Gamma_B(B)$ exhibits a broad hump with step-like structures, whereas similar features are observed in $C_p(B)$. As discussed in the previous section, the $dM/dB$ curves also show a broad anomaly around 4~T. The common appearance of these features across multiple thermodynamic probes, as well as the nonmonotonic behavior of some of the magnon modes~\cite{Pilch140406,Bischof104406} and thermal conductivity~\cite{Hong2024} across phase I, suggest that these features reflect an intrinsic response of the system. Similar low-field features in $\Gamma_B(B)$ of $\alpha$-RuCl$_3$ were attributed to the domain re-population of the single-$q$ zigzag order, but such a process should not occur in NCTO assuming the triple-$q$ magnetic order within phase I, although there may be reorientation effects associated with spin vorticity of the triple-$q$ state~\cite{Jin136701}. Magneto-optical measurements as a function of applied field could shed further light on this issue. 
		
The  increasing $\leftrightarrow$  decreasing field-sweep $\Gamma_B(B)$ data measured at $T = 2$~K are presented in Fig.~S4. In agreement with the $M(B)$ results, a clear hysteresis is observed around $B_{c1}$, consistent with the first-order nature of this transition. By contrast, no detectable hysteresis is found at $B_{c2}$ and $B_{c3}$, supporting their second-order character.
		
		To further examine the nature of these field-induced phases, we combine $C_p$ and $\Gamma_B$ to estimate the field evolution of the entropy using
		\[
		\Delta S(B) = -\int_{0}^{B} C_p(B')\,\Gamma_B(B')\, dB'.
		\]
		The variation of $\Delta S$ versus magnetic field is shown in Fig.~\ref{Fig3}(c). Since $C_p$ is always positive, a sign change in $\Gamma_B$ corresponds to an extremum in $\Delta S$. Accordingly, the sharp sign changes from negative to positive in $\Gamma_{B}$ at $B_{c2}$ and $B_{c3}$ indicate sharp local maxima in $\Delta S$ corresponding to entropy accumulation at continuous phase transitions. By contrast, the anomaly at $B_{c1}$ produces only a weaker and more broadened hump in $\Delta S$, as is often found near first-order transitions. While $C_p(B)$ and $\Gamma_{B}$ may indicate a further shoulder like anomaly between $B_{c2}$ and $B_{c3}$, the absence of a distinct maximum in the corresponding $\Delta S(B)$ data suggests that this feature is not associated with an additional phase transition. The $\Gamma_{B}$ measurements are extended to higher temperatures, as shown in Fig.~\ref{Fig3}(d). The sign change at $B_{c2}$ remains visible at all measured temperatures, while the signature associated with $B_{c3}$ gradually broadens and appears as a broad maximum near $B_{c3}$ at 4~K. This behavior is consistent with the signature of the phase transition observed in the second derivative $d^{2}M/dB^{2}$ data shown in Fig.~\ref{Fig1}(e), which also weakens with increasing temperature.
			
\begin{figure}[!ht]
		\centering
		\includegraphics[width=1\linewidth]{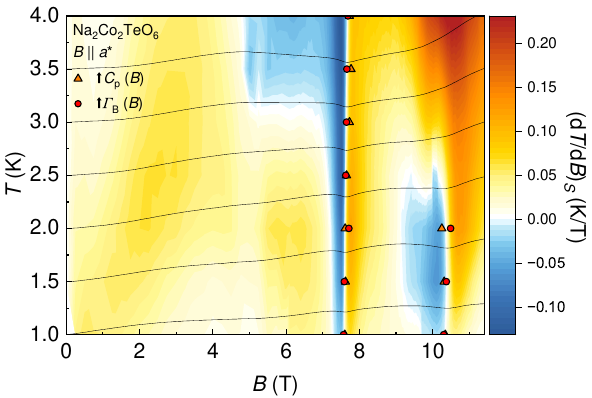}
		\caption{\sffamily\textbf{Colour map}: Colour map of the adiabatic MCE as a function of temperature $T$ and magnetic field $B$ for $B \parallel a^{*}$. The color scale represents $(dT/dB)_S$ in units of K/T, where positive (yellow--red regions) and negative (blue regions) values highlight distinct field-induced regimes. The black lines display adiabatic
			temperature traces. Pronounced features near $\sim 8$~T and $\sim 10$--$11$~T mark phase boundaries consistent with the critical fields discussed in the text. The triangles and circles are from the position of broad maxima in  $C_p(B)$ and zero-crossings in $\Gamma_{B}$ which signal the appearance and disappearance of phase III as a function of field.}
		\label{Fig4}
	\end{figure}

\begin{figure}[!ht]
		\centering
		\includegraphics[width=1\linewidth]{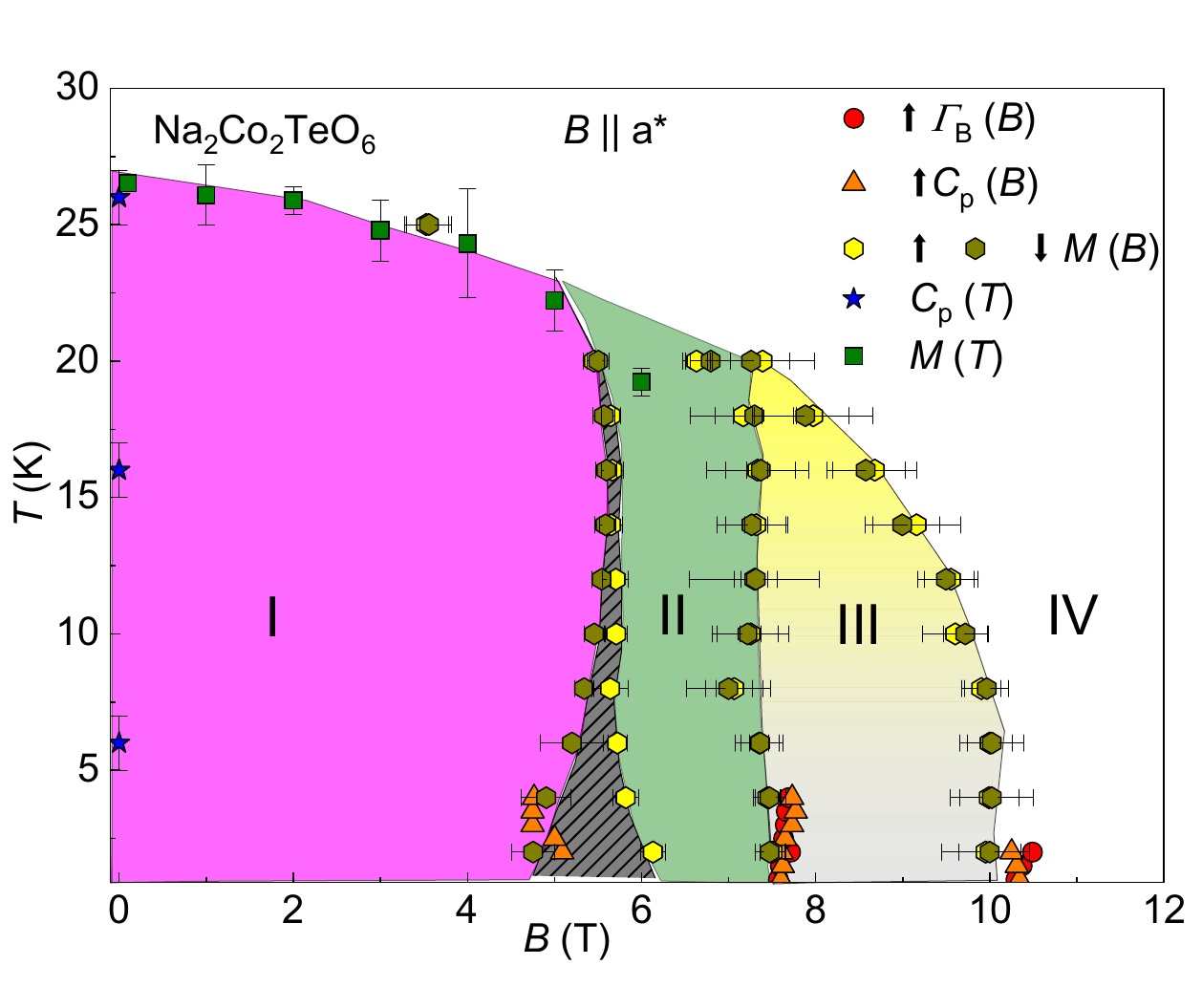}
		\caption{\sffamily\textbf{B--T phase diagram of NCTO}: To construct the magnetic phase diagram, transition points were identified from characteristic features in thermodynamic and magnetic measurements. For temperature sweeps, peaks in $d[\chi(T)T]/dT$ and anomalies in the specific heat $C_p(T)$ were used. For field sweeps, the phase boundaries were determined from peaks and broad minima in $dM/dB$ and $d^2M/dB^2$, maxima in the field-dependent specific heat $C_p(B)$, and zero crossings in the magnetic Gr\"uneisen parameter $\Gamma_B$.}
		\label{Fig5}
	\end{figure}

		Using the combined $\Gamma_B$ data, we also determine isentropes in the $B$--$T$ plane by numerical integration as shown in Fig.~\ref{Fig4}. These adiabatic temperature traces are plotted as black lines. The corresponding magnetocaloric response reveals cooling anomalies at the phase boundaries, visible as blue regions where $(dT/dB)_S < 0$, while red regions mark field ranges where the sample temperature increases under quasi-adiabatic conditions. Altogether these results establish a consistent thermodynamic picture of successive field-induced transitions in NCTO.\\
	    
	\vspace{5pt}
	
	\vspace{5pt}
	
	\noindent \textbf{Discussion}\\
	The combined field--temperature phase diagram constructed from magnetization, specific-heat, and $\Gamma_B$ results, is summarized in Fig.~\ref{Fig5}. In the low-temperature limit, the phase boundaries in Fig.~\ref{Fig5} are well captured by three critical fields, $B_{c1} \simeq 5.8$~T, $B_{c2}\simeq 7.8$~T, and $B_{c3}\simeq 10.4$~T, which partition the diagram into four regions (Phase I--IV).  
	Owing to the strong field-induced broadening of the anomalies associated with $T_{N2}$ and $T_{N3}$, we do not include them as separate lines in the phase diagram. In line with the main scope of this work, we focus on the low-temperature high-field part of the phase diagram, where a putative QSL regime has been previously discussed.

	Previously, phase III was sometimes interpreted as the QSL regime in light of an excitation continuum observed by inelastic neutron scattering~\cite{Lin2024}. However, our thermodynamic measurements provide evidence that warrants a more cautious interpretation. Notably, $C_p(B)$ exhibits clear anomalies at both the entrance to and exit from phase III, showing that this field range is bounded by well-defined thermodynamic phase transitions. Correspondingly, the magnetic entropy ($\Delta S$) shows no enhancement in this regime; instead, its evolution is opposite to what one would anticipate in a QSL state. This behavior is further corroborated by the magnetocaloric response $\mathrm{d}T/\mathrm{d}B$, which reveals sharp cooling anomalies at $B_{c2}$ and $B_{c3}$ , while the sample temperature rises monotonically between these two critical fields---a behavior more consistent with an ordered phase~\cite{Gegenwart2016}. We note, however, that Zhang \textit{et al.}~\cite{Zhang064421} pointed out that a gapped QSL would also be expected to show a decrease in magnetic entropy upon entering the phase, accompanied by an increase in lattice entropy, qualitatively consistent with the temperature rise we observe in phase III. While we cannot entirely exclude this scenario on the basis of our $\Gamma_{B}$ data alone, the presence of a sharp $C_p$ anomaly at the phase boundaries suggests that the transitions bounding phase III are thermodynamically well-defined, in contrast to a mere crossover separating a QSL from the paramagnetic state. 
	
	Interestingly, neutron diffraction studies have shown the strong suppression of the magnetic Bragg peaks at the $M$ point already above $B_{c1}$ for $B \parallel a^*$~\cite{Bera214419}, whereas the Bragg peaks persist at least up to 10~T for $B \parallel a$~\cite{YaoL022045}, but no diffraction experiments have been reported within phase III ($B \parallel a^*$) so far. THz spectroscopy measurements in the same field range reveal a large number of magnon modes~\cite{Xiang076701,Shi184406} that would be consistent with a complex but ordered state.

  	Beyond $B_{c3}$ (above $\sim 10.4$~T), none of our probes reveal any additional phase transitions, with the magnetization saturating smoothly, whereas both $C_p(B)$ and $\Gamma_{B}$ become featureless. This stands in contrast to the behavior reported in the Kitaev candidate $\alpha$-RuCl$_3$, where a shoulder-like anomaly in the magnetic Gr\"{u}neisen parameter $\Gamma_B$ persists well above the critical field, accompanied by a modest reduction in the magnetic entropy~\cite{Bachus097203}. Theoretically, this feature is attributed to a level crossing between the lowest excited states, which causes an abrupt change in the field dependence of the excitation gap and thereby produces a discontinuity in $\Gamma_B$. While the overall suppression of magnetic order with increasing field in NCTO bears a qualitative resemblance to the behavior in $\alpha$-RuCl$_3$, the high-field response of the two systems differs. In NCTO, we observe neither a $\Gamma_B$ anomaly nor any detectable entropy reduction beyond $B_{c3}$, suggesting that the system enters a conventional spin-polarized state without passing through any intermediate field-induced disordered phase. We note that a sign change in $\kappa_{xy}$ has been reported above $B_{c3}$~\cite{Guang024419}, which may hint at subtle changes in the magnetic excitation spectrum. On the other hand, spectroscopic probes suggest a continuous evolution across this field range~\cite{Xiang076701,Pilch140406}.

	\vspace{0.1cm}
		
	\begin{center}
		\textbf{METHODS}
	\end{center}
	\textbf{Crystal Growth} \\
	\noindent High-quality single crystals of NCTO were  grown using a self-flux method. A polycrystalline precursor, synthesized from Na$_2$O, Co$_3$O$_4$, and TeO$_2$, was heated to 900~$^\circ$C, held for 30~h, and then slowly cooled to 500~$^\circ$C. This controlled cooling produced light-purple, flaky crystals with typical dimensions of 1–5 mm.
	
	\vspace{0.1cm}
    \noindent \textbf{Magnetization and Specific Heat}\\
	\noindent Magnetization and specific-heat measurements above $T = 2\,\mathrm{K}$ 
	were performed in a commercial Quantum Design (QD) 14-T Physical Property 
	Measurement System (PPMS). 
	
	\vspace{0.1cm}
	\noindent \textbf{Magnetocaloric Effect and Field-dependent Specific Heat}\\
	\noindent Field-dependent measurements of the specific heat and magnetic Gr\"uneisen parameter $\Gamma_B$ were carried out in a dilution refrigerator, employing the heat pulse method for the specific heat and a high-resolution alternating-field technique for $\Gamma_B$~\cite{Tokiwa013905}.
	The measurement setup is shown in Fig.~S5 in the Supplementary figures. The magnetic field was applied along the $a^{*}$ direction and the sample temperature was monitored using a calibrated RuO$_2$ thermometer. The specific heat was measured by applying a small heat pulse \(P\) to the sample and recording the resulting temperature change. The temperature-time curves are fitted with equation~(\ref{eq:Heatpuls_fit}) to obtain the temperature increase \(\Delta T\) and relaxation time \(\tau\), which are then used to determine the heat capacity \(C_p\) via equation (\ref{eq:heat_capacity}).
	\begin{equation}
		\label{eq:Heatpuls_fit}
		T(t) = T_{\mathrm{final}} + \Delta T \cdot e^{-\frac{t}{\tau}}
	\end{equation} 
	
	\begin{equation}
		C_p = P \frac{\tau}{\Delta T}
		\label{eq:heat_capacity}
	\end{equation}

	The magnetic Grüneisen parameter was determined from the field dependence of the sample temperature under quasi-adiabatic conditions. To this end, an oscillating field with the frequency $f$ of 0.5~Hz and an amplitude \(\Delta B\) between 6~mT and 20~mT, was superposed onto the dc magnetic field. Fitting of the field and temperature oscillations with equations~(\ref{eq:field_oscillation}) and (\ref{eq:temperature_oscillation}), allows one to determine the amplitudes for the magnetic field and temperature (\(\Delta B\)  and \(\Delta T\)). Now, using equation~(\ref{eq:Grüneisenparameter}) the magnetic Gr\"uneisen parameter \(\Gamma_B\) is determined by using previously obtained \(\Delta B\) and \(\Delta T\). The term \(\delta t\) in equations~(\ref{eq:field_oscillation}) and ~(\ref{eq:temperature_oscillation}) is a small delay due to a finite thermal conductivity between the sample and thermometer~\cite{Tokiwa013905}. The background term \(b_0 + b_1 t + b_2 t^2\) in equation~(\ref{eq:temperature_oscillation}) is included to account for possible temperature drift.
	
		\begin{equation}
		B(t) = \Delta B \cdot \sin{\left(2\pi f(t-\delta t )\right)} + B_0
		\label{eq:field_oscillation}
	\end{equation}
	
	\begin{equation}
		\begin{split}
			T(t) &= \Delta T \cdot \sin\left(2\pi f (t-\delta t)\right) \\
			& \quad + \Delta T_{2f} \cdot \cos\left(4\pi f (t-\delta t)\right) \\
			& \quad + b_0 + b_1 t + b_2 t^2
		\end{split}
		\label{eq:temperature_oscillation}
	\end{equation}
	
	\begin{equation}
		\Gamma _B = \frac{1}{T} \left(\frac{\partial T}{\partial B}\right)_S
		\label{eq:Grüneisenparameter}
	\end{equation}
	
	To calculate the entropy evolution with respect to the magnetic field, we use the relation in equation~(\ref{eq:entropie}), which is derived from the Maxwell relation \(\left(\frac{\partial S}{\partial B}\right)_T = \left(\frac{\partial M}{\partial T}\right)_p\) and the definition of the magnetic Gr\"uneisen parameter \(\Gamma_B = -\frac{1}{C_p} \left(\frac{\partial M}{\partial T}\right)_B\).
	
	\begin{equation}
		\Delta S(B) = -\int_{0}^{B} C_p(B')\,\Gamma_B(B')\, dB'
		\label{eq:entropie}
	\end{equation}
	
	\begin{center}
		\textbf{REFERENCES}
	\end{center}
		
	\bibliography{bib}
		
	\vspace{0.2cm}
		
	\noindent\textbf{DATA AVAILABILITY}\\
	 The data asscociated with the main results of this manuscript can be found in the Zenodo online repository with unique identifier: {\color{blue} 10.5281/zenodo.19597936}.\\
		
	\noindent\textbf{ACKNOWLEDGEMENTS}\\
     We thank Lukas Janssen and Yuan Li for fruitful discussions. This work was funded by the Deutsche Forschungsgemeinschaft (DFG, German Research Foundation) -- TRR 360 -- 492547816 (subproject B1). The work at SKKU was supported by the National Research Foundation (NRF) of Korea (Grants No. RS-2023-00209121 and No. 2020R1A5A1016518). R.S. acknowledges the financial support provided by the Ministry of Science and Technology in Taiwan under Projects No. NSTC-114-2124-M-001-009 and NSTC-113-2112M001045-MY3, as well as support from Academia Sinica for the budget of AS-iMATE11514.
	\vspace{0.2cm}
		
	\noindent\textbf{AUTHOR CONTRIBUTIONS}
		
	\noindent P.G., A.-A.T and K. Y. C. designed the project. P. K. M. and S. E. performed the magnetization and specific heat measurements above 2\,K. S. E. carried out and analyzed the MCE and specific heat measurements in the dilution refrigerator. R.K. and R.S. grew the single crystals. All authors discussed the results. P.K.M, S.E., P.G. and A.-A.T wrote the manuscript.
	\vspace{0.2cm}
		
	\noindent\textbf{COMPETING INTERESTS}\\
	The authors declare no competing interests.
	
\end{document}